\documentclass[]{spie}  

\usepackage{amsmath,amsfonts,amssymb}
\usepackage{graphicx}
\usepackage[colorlinks=true, allcolors=blue]{hyperref}
\usepackage{cleveref}
\usepackage{siunitx}
\newcommand{\capleft}{\emph{(Left)}}
\newcommand{\capright}{\emph{(Right)}}

\newcommand{\capmiddle}{\emph{(Middle)}}
\newcommand{\capcenter}{\emph{(Center)}}
\newcommand{\captext}[1]{\emph{(#1)}}

\title{Characterization of Mid-Infrared Intersubband Detectors for Astronomical Heterodyne Interferometry}

\author{Tituan Allain\supit{a}\supit{b}, Mohammadreza Saemian\supit{b} Carlo Sirtori\supit{b},Jean-Philippe Berger\supit{a} \skiplinehalf \supit{a}Univ. Grenoble Alpes, CNRS, IPAG, 38000 Grenoble, France \\ \supit{b} Laboratoire de Physique de l’Ecole Normale Supérieure, ENS, Université PSL, CNRS, Sorbonne Université, Université de Paris, 24 rue Lhomond, 75005, Paris, France}

\authorinfo{Further author information: (Send correspondence to T.A.)\\T.A.: E-mail: tituan.allain@univ-grenoble-alpes.fr\\  J-P.B.: E-mail: jean-philippe.berger@univ-grenoble-alpes.fr}

\begin{document} 
\maketitle

\begin{abstract}
One of the major challenges of mid-infrared astronomical heterodyne interferometry is its sensitivity limitations. Detectors capable of handling several \SI{10}{\giga\hertz} bandwidths have been identified as key building blocks of future instruments. Intersubband detectors based on heterostructures have recently demonstrated their ability to provide such performances. In this work we characterize a Quantum Well Infrared Photodetector in terms of noise, dynamic range and bandwidth in a non-interferometric heterodyne set-up. We discuss the possibility to use them on astronomical systems to measure the beating between the local oscillator and the astronomical signal.

\end{abstract}


\section{INTRODUCTION}

The study of planet formation and protoplanetary disk evolution around young star requires high resolution angular imaging and spectroscopy. Angular resolutions of less than $\SI{1}{mas}$ are necessary to distinguish or isolate the main features and structures of these objects\cite{monnier2018}. The emission spectra of dust and young planets peak in the mid-infrared. In the astronomical N band ($[8,12]\si{\micro\meter}$), $\SI{1}{mas}$ corresponds to kilometric baselines. In addition, a large number of baselines is required to resolve complex structures in protoplanetary disk. Thus, astronomical interferometry with many telescopes, typically tens of telescopes\cite{monnier2018}, would be the only possibility to obtain these imaging resolutions. Direct mid-infrared interderometer already exist, like the VLTI which can be set up to $\SI{130}{\meter}$ baseline with four telescope. This type of interferometer directly combine the light from an array of telescope, either two by two or altogether. This would hardly be scalable to a large number of telescope, mostly because of the bulkiness of the delay-lines that are required for the beam combination. Heterodyne interferometry would be an alternative to the direct interferometry approach but suffers from reduced sensitivty\cite{ireland2014}.

\begin{figure}[t]
    \centering
    \includegraphics[width=0.8\linewidth]{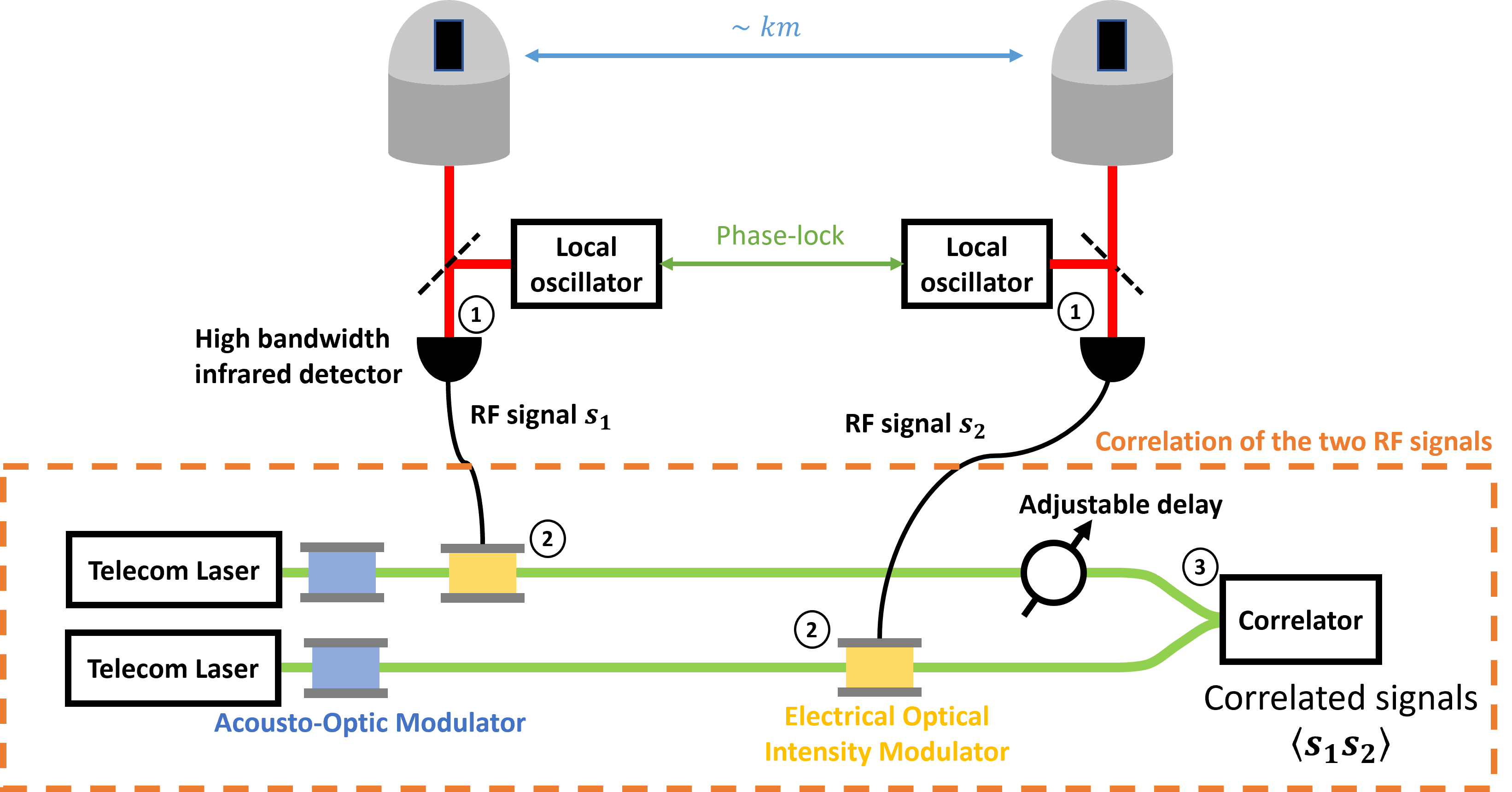}
    \caption{Principle of two telescope astronomical heterodyne interferometry.}
    \label{fig:heterodyne_interferometry}
\end{figure}

The basic principle of heterodyne interferometry is ilustrated by \cref{fig:heterodyne_interferometry} on a typical two telescope set-up. First, the signal from the astronomical object at frequency $\nu_s$ is collected by the two telescopes. It is mixed with a local oscillator at frequency $\nu_{\text{LO}}$ at the focal point of each telescope, thus creating radio-frequency signals $s_1$ and $s_2$ at frequencies $\Delta\nu=|\nu_s-\nu_{\text{LO}}|$. The heterodyne beatings preserve the phase information of the astronomical signal. The relative phase between the local oscillator at each telescope must be stabilized for the interferometric phase information to be retrieved. Second, the two RF signals are encoded onto an optical fibre telecom signal carrier, enabling long distance travel and signal amplification without bulky infrastructures. Third, the two signal are correlated and the interferometric observables can be retrieved. Signal correlation can done analogically using a correlation loop that can provide correlation values at multiple time delays\cite{bourdarot2022}. A compensation delay can be added before the correlator to one of the signal to compensate for baseline delay between the two telescopes. The sensivity of an heterodyne interferometer is driven by the signal to noise ratio $SNR_{\text{heterodyne}}$ of the heterodyne detection itself, given by \cref{eq:SNR} in the shot-noise limited regime.

\begin{equation}
    SNR_{\text{heterodyne}}\propto\frac{\text{d}\Phi_{s}(\nu)}{\text{d}\nu}\frac{\sqrt{2\Delta\nu t}}{h\nu}
    \label{eq:SNR}
\end{equation}

Where:
\begin{itemize}
    \item $\frac{\text{d}\Phi_{s}(\nu)}{\text{d}\nu}$ is the spectral power of the astronomical signal, expressed in $[\si{\watt/\hertz}]$
    \item $\nu$ is the frequency of the signal
    \item $h$ is the Planck constant
    \item $\Delta \nu$ is the bandwidth of the detector
    \item $t$ is the integration time of the heterodyne signal
\end{itemize}

In the ideal case of a shot-noise limited detector, $SNR_{\text{heterodyne}}$ scales as the square root of the detector bandwidth. Thus, a mid-infrared astronomical heterodyne interferometer should maximize the detection bandwidth while keeping the noise level of the detector down to ensure a good $SNR$. In the recent years, a new generation of mid-infrared detector has emerged, namely intersubband detectors. They rely on the quantum confinement of electrons in the conduction band of a semi-conductor. These detectors exhibit bandwidth of several tens of $\si{\giga\hertz}$ with high specific detectivty \cite{palaferri2018}. The low-noise even enable the use of these detectors at ambient temperature. These detectors are categorized in two main categories: Quantum Cascade Detectors (QCD) and Quantum Well Infrared Photodetectors (QWIP). In this paper, we present intersubband detectors in the scope of astronomical application for heterodyne interferometry. In the first part, we describe the basic principle of intersubband detectors, QCDs, QWIPs and the use of meta-material for improved performances. In the second part, we describe the experimental methods that can be used to characterize these detectors in terms of responsivity, noise level, specific detectivity, bandwidth and  we apply these methods on a $\SI{5}{\micro\meter}$ QWIP.

\section{PRINCIPLE OF QUANTUM CASCADE AND QUANTUM WELL INFRARED DETECTORS}
In this part, we first describe the basic physics of intersubband transitions which are used we both QCD and QWIP. Then we describe the principle of these two types of detectors. Finally we describe the mesa and meta-material based patch-antenna architectures that are used to implement QWIPs and QCDs.

\subsection{Quantum wells and intersubband transitions}
Quantum infrared detectors rely on electrons being promoted from a ground state to an excited state by photon absorption. In the well-known interband detectors, like classical photodiodes, the electrons are promoted from the valence band to the conduction band that are represented on \cref{fig:patch_antenna} \capleft. The energy gap of the semiconductor defines the cut-off frequency of the detector and drastically limits the types of semiconductors that can be used for classical photodetection in mid infrared. This limit can be overcome by using heterostructures and intersubband transitions inside the conduction band. Different layers of semiconductors are superimposed to confine the electrons inside quantum wells as shown in \cref{fig:patch_antenna} \capcenter. The thickness $l$ of the quantum well defines the eigen-energies levels $E_n$ inside the quantum well. For an infinite potential barrier, the energy level are given by \cref{eq:eigen_energies} where $m^*$ is the effective mass of the electron. The intersubband transition occurs between states $E_1$ and $E_2$ of the quantum well. The quantum wells are n-doped to increase the Fermi level and make sure the ground state of the quantum well is populated by electrons. Unlike interband based detectors which use electron-holes pairs, intersubband based detectors only use electrons: they are called unipolar devices. The main advantage of a unipolar device comes from the higher mobility of the electrons and thus from the possibility to use fast scattering and tunnelling processes which occur at the $\sim \SI{10}{\pico\second}$ timescale \cite{steinkogler2003}.

\begin{equation}
    E_n=\frac{\hbar^2\pi^2}{2m^*l^2}n^2
    \label{eq:eigen_energies}
\end{equation}

The very fast physical phenomena behind absorption, scattering and tunneling in both QCD and QWIP enable very high detection bandwidth which could, in theory, go as far as $1/\SI{10}{\pico\second}=\SI{100}{\giga\hertz}$. Bandwidth of tens of $\si{\giga\hertz}$ have already been achieved\cite{hakl2021}. The main limitation for reaching higher bandwidths is the capacitance of the electrical structure of the detector. Under this limitation, the bandwidth a detector is proportional to the inverse of its electrical area. 

\subsection{Quantum Well Infrared Photodetectors}

\begin{figure}[]
    \centering
    \includegraphics[width=1\linewidth]{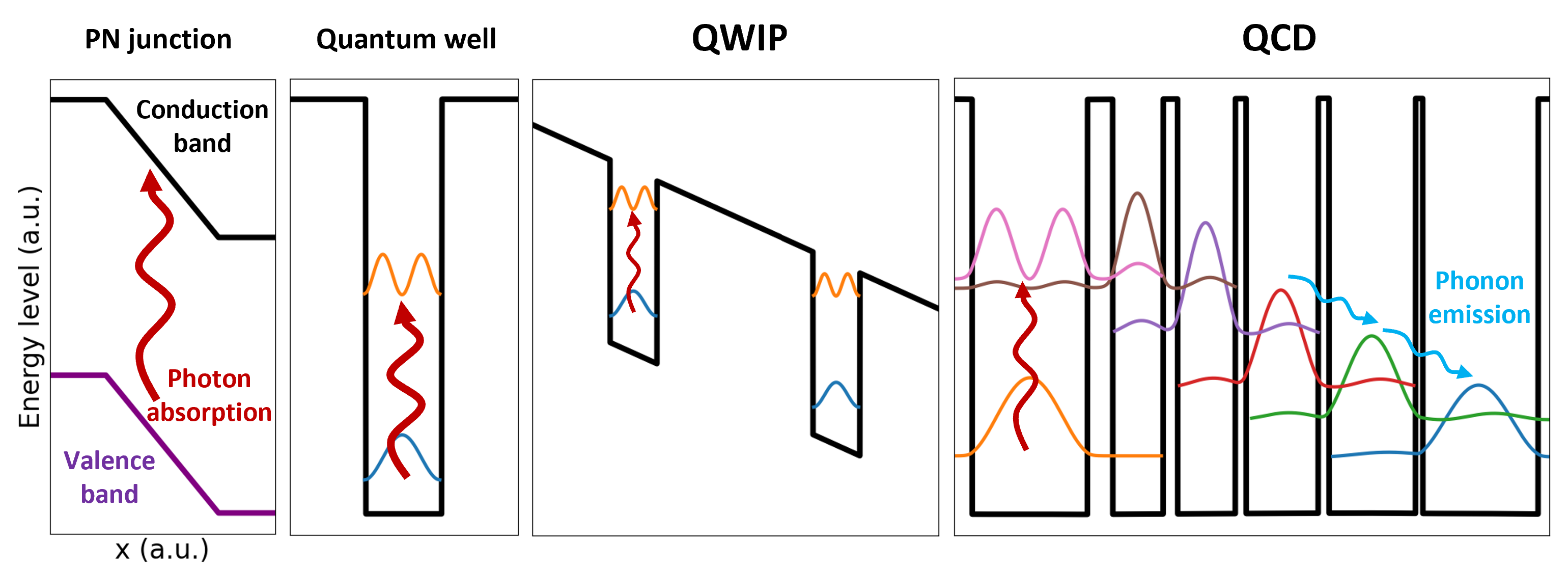}
    \caption{Energy levels inside different structures. From left to right: a semiconductor, a single quantum well, a QWIP and a QCD. For the three last, only the conduction band is represented.} 
    \label{fig:QCD_QWIP}
\end{figure}

\begin{figure}[b]
    \centering
    \includegraphics[height=5cm]{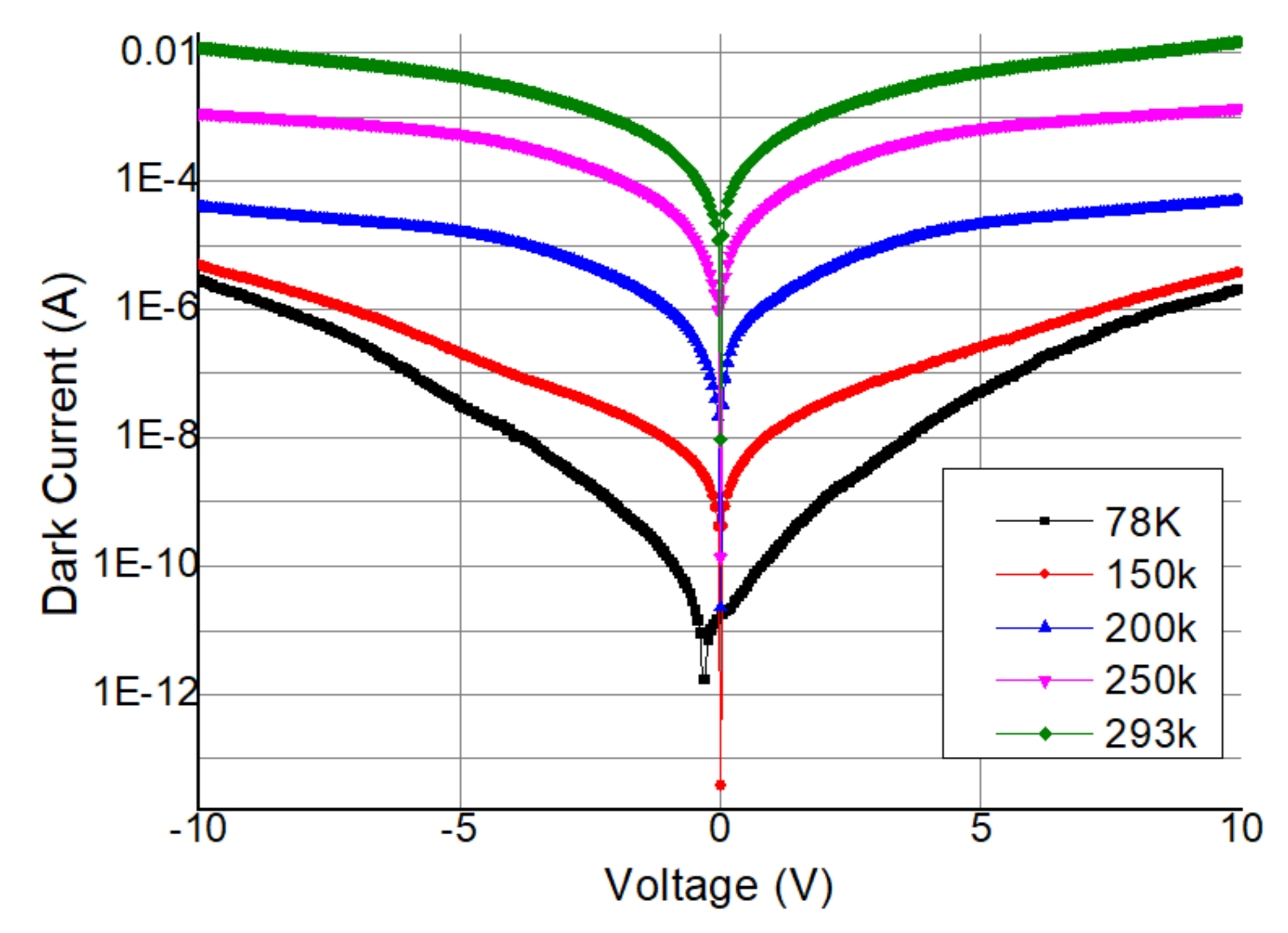}
    \caption{Dark current of a $\SI{5}{\micro\meter}$ QWIP as a function of applied bias and operating temperature.}
    \label{fig:qwip_dark_current}
\end{figure}

Quantum Well Infrared Photodetector consist in a series of identical quantum well separated by thick barriers. The diagram of a QWIP is given by \cref{fig:QCD_QWIP}. The quantum well are designed so that the excited $E_2$ states are close to barrier tops. The electrons that are promoted to the excited states and are extracted by applying an external voltage bias which breaks the symmetry of the QWIP structure and imposes a flowing direction for the electrons. The higher the bias, the higher the responsivity of the detector. However, the bias also generates dark current due to thermal activation that promotes electrons to the continuum. The more bias, the higher the dark current. Decreasing the operating temperature of the QWIP curbs dark current and its associated noise. QWIPs are also subject to the Johnson-Nyquist noise which is caused the thermal fluctuation of the current and also decreases with lower operating temperature. In \cref{fig:qwip_dark_current}, the dark-current of a $\SI{5}{\micro\meter}$ QWIP is shown as a function of applied bias for different operating temperatures. The lower temperature, the less dark-current. At $\SI{-5}{\volt}$ bias, the dark current is reduced from $\SI{4}{\milli\ampere}$ at $\SI{293}{\kelvin}$ down to $\SI{30}{\nano\ampere}$ at $\SI{78}{\kelvin}$. 

\subsection{Quantum Cascade Detectors}
Quantum Cascade Detectors consist in a periodic series of quantum well that are designed to extract the electrons via tunneling processes without applying any bias. The asymmetric structure of the QCD drives the flowing direction for the electrons. The typical layer structure of a QCD is depicted in \cref{fig:QCD_QWIP}. The $n$\emph{-th} energy level inside a quantum well $x$ is written $E_n^x$. First, electrons from main doped quantum wells are promoted by photon absorption. The electrons are extracted by resonant tunneling with adjacent quantum wells. The $E_1^b$ energy level of the first adjacent quantum well is made similar to the energy level $E_2^a$ of the promoted electron in the main quantum well, thus creating a strong coupling between the two levels. Then, level $E_1^c$ from the next quantum well is made so that $E_2^a=E_2^c+E_{phonon}$ where $E_{phonon}$ is the typical phonon energy of the lattice. By emitting a phonon, electrons tunnel through the potential barrier one by one until the reach the end of the so-called quantum cascade. In the absence of bias, no dark current flows through a QCD detector which highly reduces the noise level of the detector, at the cost of a lower responsitivity. QCD are still subject to Johnson-Nyquist noise due to thermal current fluctuations.

\subsection{Mesa architecture for QWIP and QCD}

The selection rules state that intersubband transitions in quantum well can only occur if the electrical field has a non-zero component along the $z$ direction which is the growth direction of the quantum wells \cite{capasso1999}. Therefore, normal incident light with respect the heterostructures will not trigger any intersubband absorption. The most commonly used architecture for QWIP and QCD is called the mesa structure and is represented on \cref{fig:QCD_QWIP}. The infrared radiation is sent on a $\SI{45}{\degree}$ GaAs facet and is then coupled into the active regions of the detector. In such a structure, polarization selection rules reduce the coupling between the infrared electrical field and the quantum well by $\sin^2(\SI{45}{\degree})=\SI{50}{\percent}$.

\subsection{Patch-antenna meta-material for QWIP and QCD}

\begin{figure}[]
    \centering
    \begin{minipage}{0.32\textwidth}
        \includegraphics[width=1\linewidth]{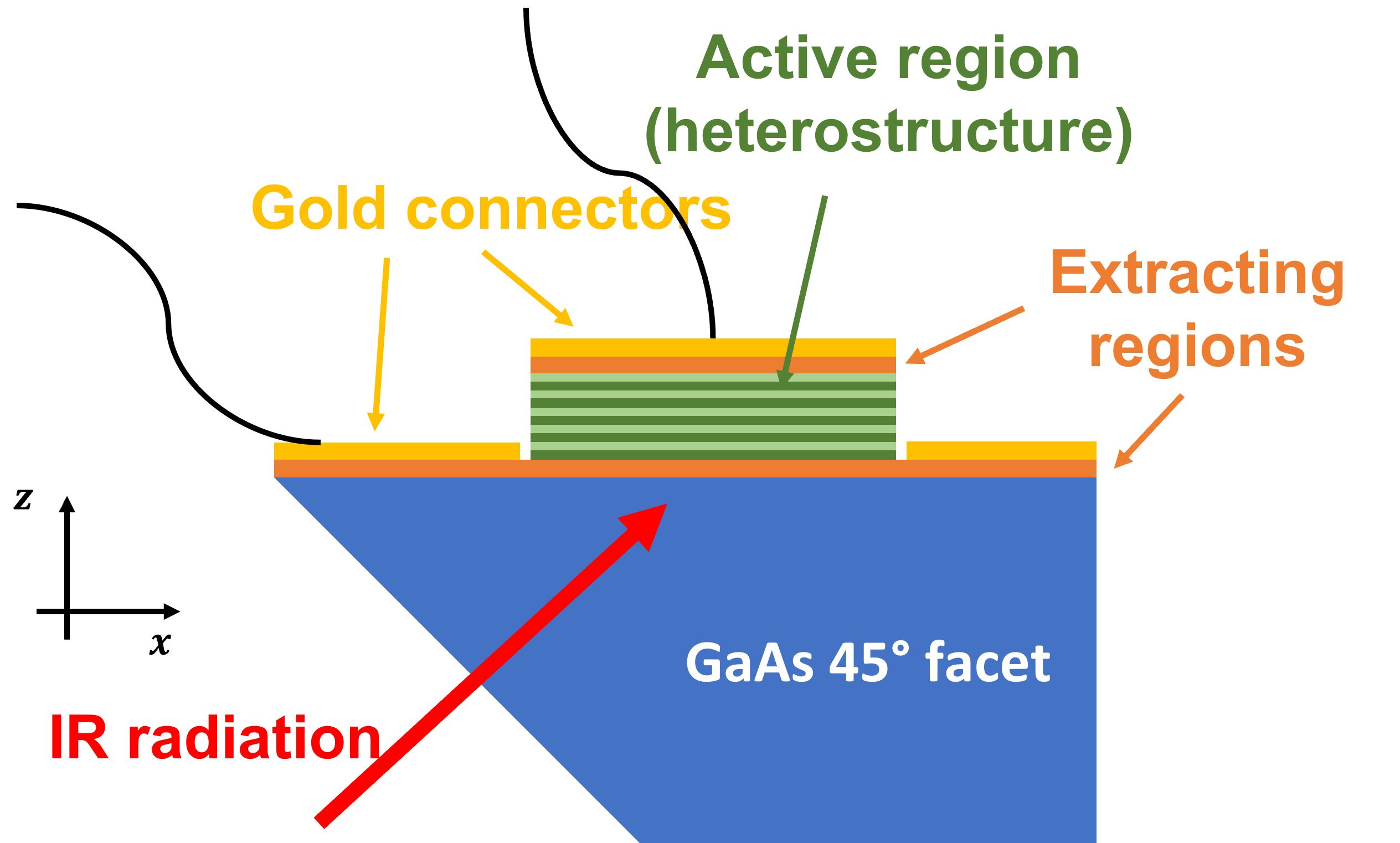}
    \end{minipage}
    \hfill
    \begin{minipage}{0.41\textwidth}
        \includegraphics[width=1\linewidth]{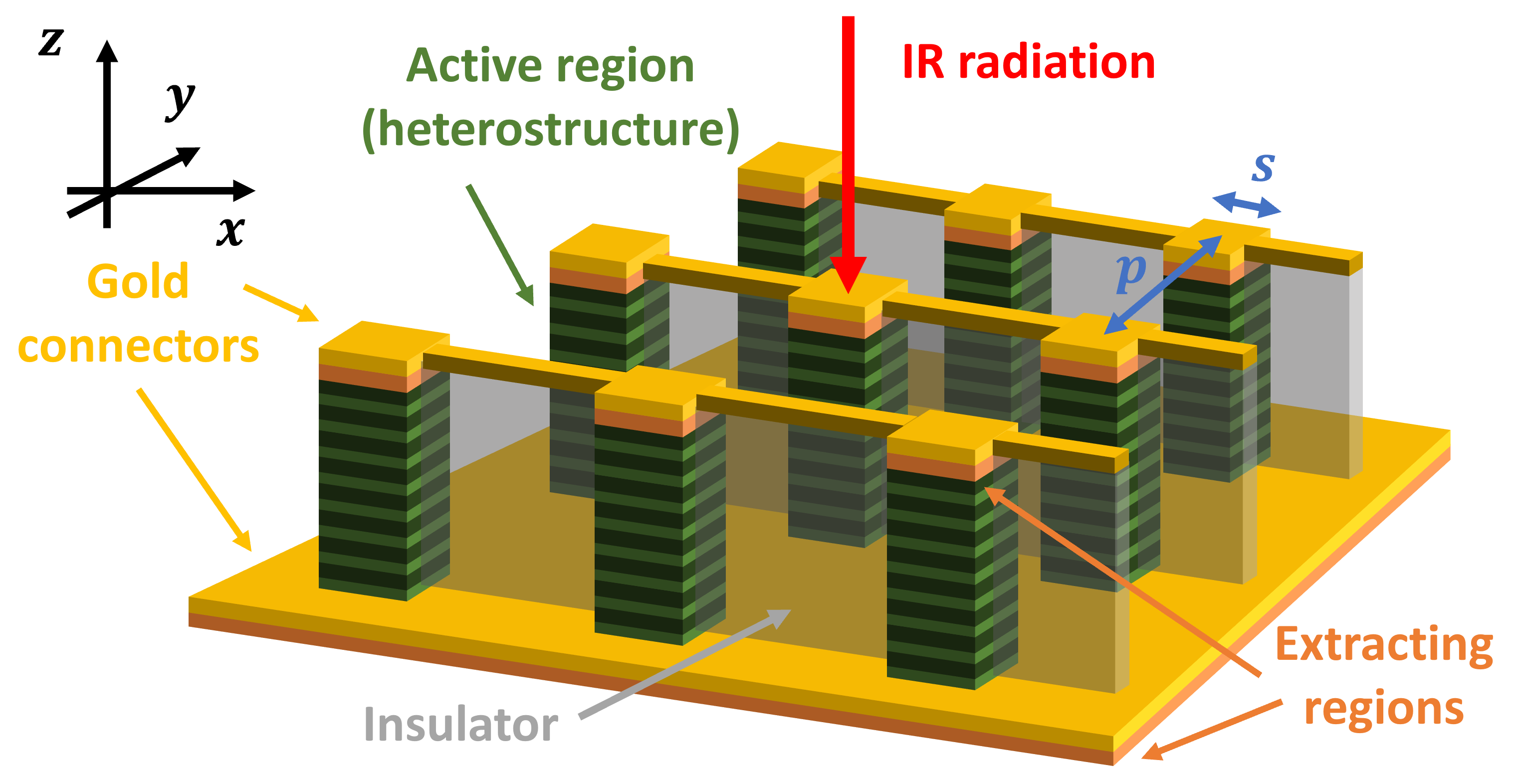}
    \end{minipage}
    \hfill
    \begin{minipage}{0.25\textwidth}
        \includegraphics[width=1\linewidth]{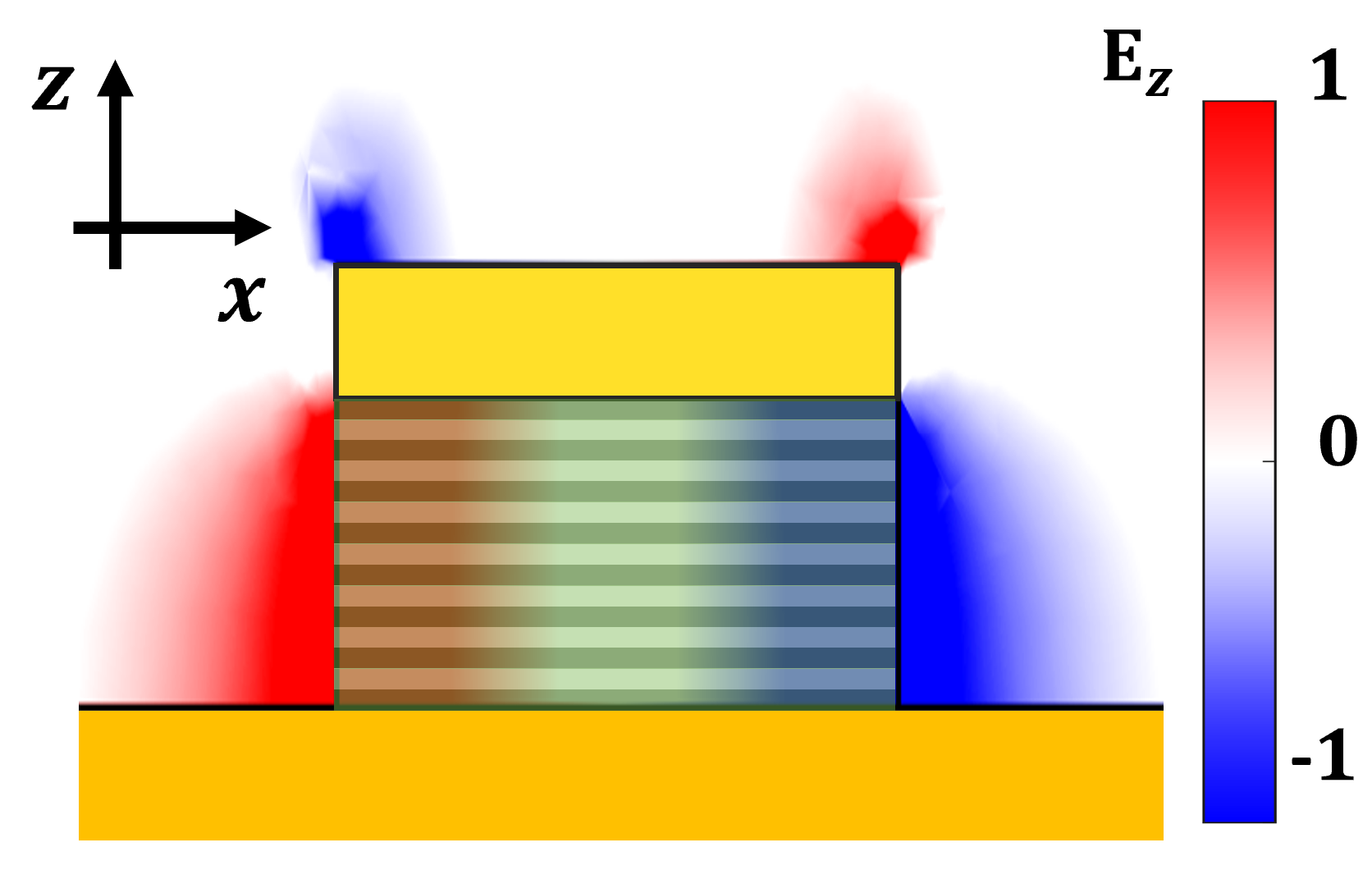}
    \end{minipage}
    \caption{\capleft~Schematics of a mesa architecture. \capmiddle~Schematics of a patch-antenna resonator.  \capright~Simulated $E_z$ component of the coupled electrical inside a single resonator under normal incident light.}
    \label{fig:patch_antenna}
\end{figure}

To overcome the intrinsic limitations of the mesa architecture, meta-material technologies can be used. In this part, we present the patch-antenna resonators which consist in a grid of sub-wavelength resonators of dimension $s\times s \times h$ made of the active region of the detector. As shown on \cref{fig:patch_antenna} \capleft, the resonators are spaced by a distance $p$ and electrically connected in lines by gold wires. Such devices are obtained using lithography and etching processes. The main advantage of this technology is its capacity to couple the incident light field inside the resonators\cite{rodriguez2022}, even with normal incident field. Thus, the theoretical amount of light that can be absorbed by the active region is not limited by the polarization selection rules. The $E_z$ component of the electrical inside a single resonator under normal incident light is shown in \cref{fig:patch_antenna}~\capright. Additionally, the electrical area of a patch-antenna detector is reduced compared with a mesa-detector, resulting in a lower capacitance which is inversely proportional to the electrical surface area, thus increasing the bandwidth of the detector. The smaller active region also generate less dark current, reducing the noise. 

\section{EXPERIMENTAL CHARACTERIZATION OF QCD AND QWIP}

In this section, we describe experimental methods to characterize QWIP and QCD detectors.  These methods can be used for any QCD or QWIP in mid-IR and can be performed at low temperature in a cryostat. We performed  measurements on a $\SI{5}{\micro\meter}$ wavelength QWIP with mesa architecture at ambient temperature. A $\SI{-9}{\volt}$ bias was applied on the detector. Our ultimate goal is to operate in the astronomical N band ($[8,12]\si{\micro\meter}$) but at the time of the writing the work was done at $\SI{5}{\micro\meter}$ in order to validate the experimental methods and provide a first look into the effect of technological parameters. The results of these measurements on individual detectors will later be used to integrate QCD and QWIP detector in a complete demonstrator for astronomical heterodyne interferometry.

\subsection{Responsivity measurement}

To measure the detector responsivity, we use a reference detector whose peak responsivity $R_{\text{peak,ref}}~[\si{\volt/\watt}]$ is known. In the following equations, the reference detector in named $\text{ref}$ while the unknown detector in named $det$. The tension $U_{LI,\text{ref}}$ is extracted from the lock-in amplifier. It can be expressed as a function of the incoming field \cref{eq:lock_in_V_ref}. The same formula applies for the unknown detector and its lock-in tension $U_{\text{LI,det}}$.

\begin{equation}
	U_{\text{LI,ref}}=A_{\text{ref}}R_{\text{peak,ref}}\Phi_{\text{bb,peak}}\int \tilde{R}_{\text{ref}}(\nu)\frac{\text{d}\tilde{\Phi}_{\text{bb}}(\nu)}{\text{d}\nu}\text{d}\nu
	\label{eq:lock_in_V_ref}
\end{equation}

Where:
\begin{itemize}
	\item $\frac{\text{d}\Phi_{bb}}{\text{d}\nu}$ is the black body spectral irradiance $\si{\watt/\meter^2/\hertz}$ that arrives on the detector. The perfect theoretical black body spectrum is given by Planck's law but in a real set-up, the spectrum will be affected by air absorption and scattering, by optical losses and by the numerical aperture of the system.
	\item $\frac{\text{d}\tilde{\Phi}_{\text{bb}}}{\text{d}\nu}$ is the  normalised spectrum with a unitary maximum of $\SI{1}{\hertz^{-1}}$ We have $\frac{\text{d}\Phi_{bb}}{d\nu}=\Phi_{\text{bb,peak}}\frac{\text{d}\tilde{\Phi}_{\text{bb}}}{\text{d}\nu}$
	\item $A_{\text{ref}}$ is the detection area of the reference detector
	\item $R_{\text{ref}}$ is the spectral responsivity of the reference detector. $\tilde{R}_{\text{ref}}$ is the normalised responsivity with a maximum of $1$. We have $R_{\text{ref}}=R_{\text{peak,ref}}\tilde{R}_{\text{ref}}$
\end{itemize}

\begin{figure}[t]
	\centering
	\begin{minipage}{0.39\textwidth}
        \includegraphics[width=1\linewidth]{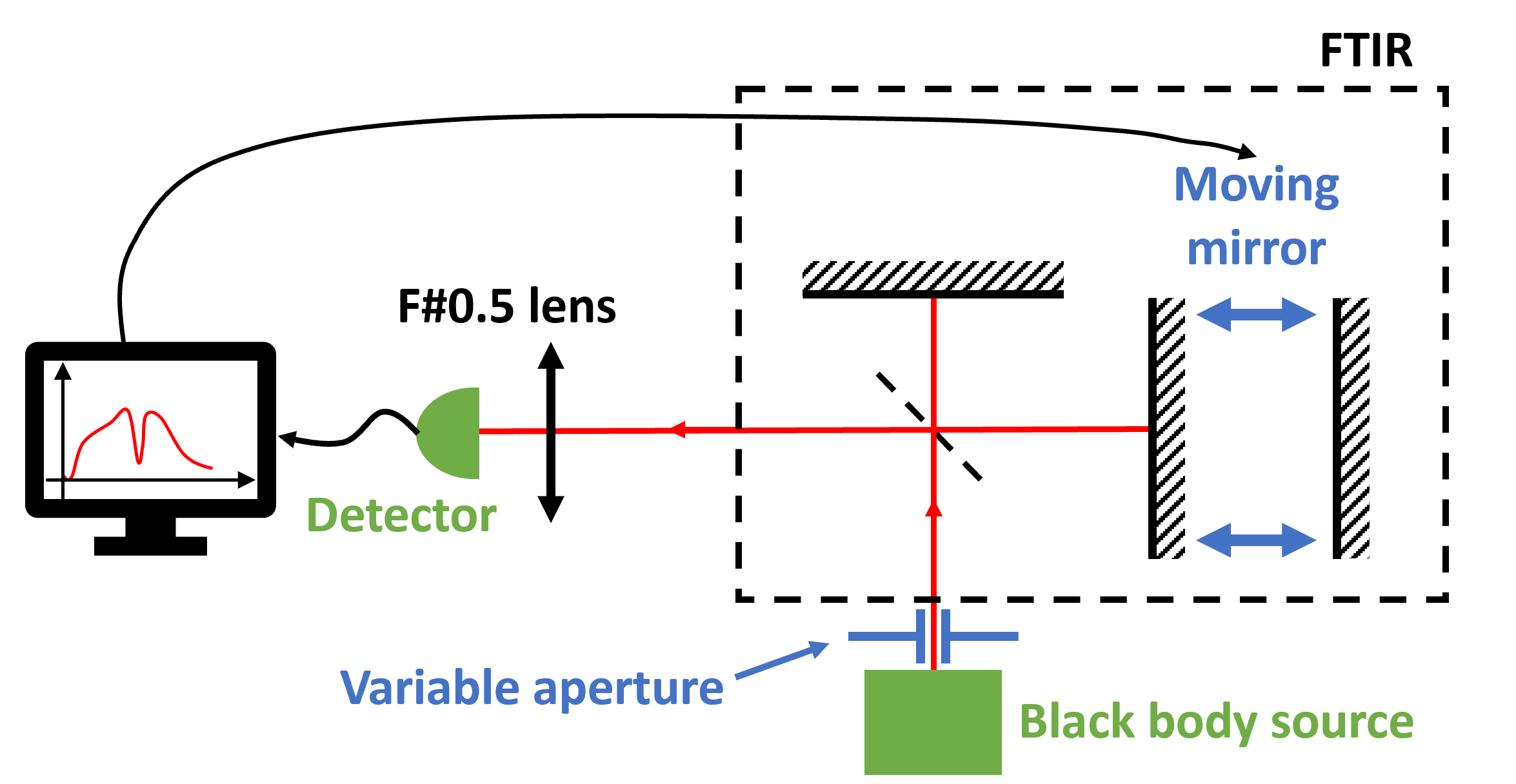}
    \end{minipage}
    \hfill
    \begin{minipage}{0.6\textwidth}
        \includegraphics[width=1\linewidth]{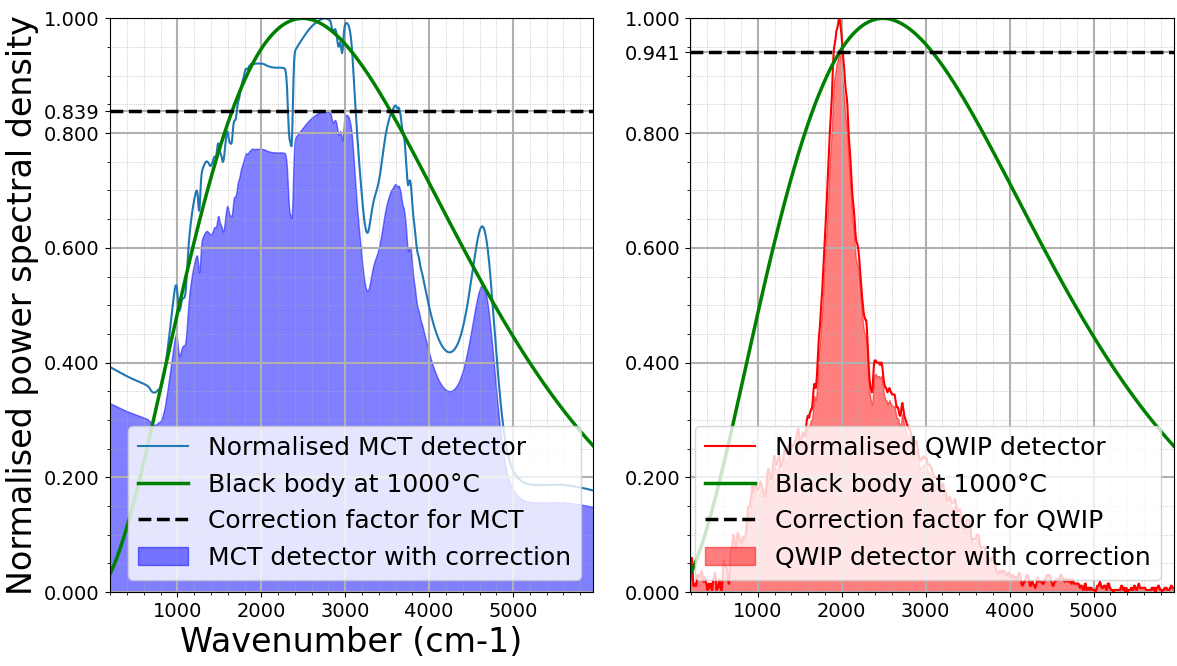}
    \end{minipage}
	\caption{\capleft~Fourier-Transform Infrared Spectrometer (FTIR) set-up for measuring the response of a detector. \captext{Middle \& Right}~Measured spectra of a $\SI{1000}{\celsius}$ black body using a reference MCT detector \capmiddle~and an uncharacterised QWIP \capright. The theoretical black body curve from Planck's law is represented in green. The blue and red surfaces respectively correspond to $\int\tilde{S}_{\text{MCT}}(\nu)d\nu$ and $\int\tilde{S}_{\text{QWIP}}(\nu)d\nu$.}
	\label{fig:MCT_QWIP_spectrums}
\end{figure}

We use the FTIR to measure the spectral response of the reference and unknown detectors with the black body source at temperature $T$. The measurement set-up is shown on \cref{fig:MCT_QWIP_spectrums}~\capleft: it uses a black body source and a Fourier-Transform Infrared Spectrometer (FTIR). A black-body source radiation is sent on a Michelson interferometer. The ouput is focused onto the detector. By varying the path difference between the two arms of the interferometer, we retrieve the temporal interferogram of the source spectrum which can be converted to the source spectrum by Fourier transform. We obtain curves corresponding to $S_{ref}(\nu)=\Gamma_{\text{ref}}\tilde{R}_{\text{ref}}(\nu)\frac{d\tilde{\Phi}_{\text{bb}}(\nu)}{d\nu}$ and $S_{det}(\nu)=\Gamma_{\text{det}}\tilde{R}_{\text{det}}(\nu)\frac{\text{d}\tilde{\Phi}_{\text{bb}}(\nu)}{\text{d}\nu}$ where $\Gamma_{\text{ref}}$ and $\Gamma_{\text{det}}$ are two coefficients that must be determined\footnote{$\Gamma_{\text{ref}}$ and $\Gamma_{\text{det}}$ can be determined knowing the frequencies $\nu_{\text{max,ref}}$ and $\nu_{\text{max,det}}$ at which the two detector have maximal responsivity. Since $\tilde{R}_{\text{det}}(\nu_{\text{max,det}})=\tilde{R}_{\text{ref}}(\nu_{\text{max,ref}})=1$, we know that $\tilde{S}_{\text{ref}}(\nu_{\text{max,ref}})=\frac{d\tilde{\Phi}_{\text{bb}}(\nu_{\text{max,ref}})}{d\nu}$ and $\tilde{S}_{\text{det}}(\nu_{\text{max,det}})=\frac{\tilde{\Phi}_{\text{bb}}(\nu_{\text{max,det}})}{d\nu}$.
The spectrums can therefore be normalised by $\tilde{S}_{\text{ref}}(\nu)=S_{ref}(\nu)/S_{ref}(\nu_{\text{max,ref}})\frac{\text{d}\tilde{\Phi}_{\text{bb}}(\nu)}{d\nu}(\nu_{\text{max,ref}})$ and $\tilde{S}_{\text{det}}(\nu)=S_{det}(\nu)/S_{det}(\nu_{\text{max,det}})\frac{\text{d}\tilde{\Phi}_{\text{bb}}(\nu)}{\text{d}\nu}(\nu_{\text{max,det}})$. The normalisation method described earlier is only valid if $\nu_{\text{max,ref}}$ and $\nu_{\text{max,det}}$ are not too much affected by air absorption, optical transmission or any kind of issue that distorts the blackbody spectrum.} to obtain the normalized spectrums $\tilde{S}_{\text{ref}}(\nu)=\tilde{R}_{\text{ref}}(\nu)\frac{\tilde{\Phi}_{\text{bb}}(\nu)}{d\nu}$ and $\tilde{S}_{\text{det}}(\nu)=\tilde{R}_{\text{det}}(\nu)\frac{\tilde{\Phi}_{\text{bb}}(\nu)}{d\nu}$. The normalized spectra that were obtained on the MCT detector and on the QWIP detector are shown on \cref{fig:MCT_QWIP_spectrums}~\captext{Middle \& Right}. The QWIP detector exhibit a narrow spectrum that is centered around $\SI{2000}{\cm^-1}$ while the MCT shows a much broader spectrum that follows the spectral shape of the incoming back-body radiation. Deviations of the MCT spectra from the Planck law are caused by losses in the system, air absorption and by the spectral responsivity of the detector.

We can calculate the optical power $P_{det} [W]$ that is received by the unknown detector using \cref{eq:P_det}. The factor $1/2$ comes from the mesa architecture of the detector which causes loses due to the $\SI{45}{\degree}$ facet. The value of $\Phi_{\text{bb,peak}}$ is given by \cref{eq:phi_bb_peak}. The total integrated response of the detector is given by $R_{\text{peak,det}}=U_{\text{0,det}}/P_{det} [\si{\volt/\watt}]$. The full expression for $R_{\text{peak,det}}$ is given by \cref{eq:R_total}.

\begin{equation}
	\Phi_{\text{bb,peak}}=\frac{U_{\text{LI,ref}}}{ A_{\text{ref}}R_{\text{peak,ref}}\int \tilde{R}_{\text{ref}}(\nu)\frac{\text{d}\tilde{\Phi}_{\text{bb}}(\nu)}{\text{d}\nu}\text{d}\nu}
	\label{eq:phi_bb_peak}
\end{equation}

\begin{equation}
	\begin{aligned}
		P_{det}=&\frac{1}{2}A_{\text{det}}\Phi_{\text{bb,peak}}\int \tilde{R}_{\text{det}}(\nu)\frac{\text{d}\tilde{\Phi}_{\text{bb}}(\nu)}{\text{d}\nu}\text{d}\nu
		\\
		=&\frac{1}{2}\frac{U_{\text{LI,ref}}}{R_{\text{peak,ref}}}\frac{A_{\text{det}}}{A_{\text{ref}}}\frac{\int \tilde{S}_{\text{det}}(\nu)\text{d}\nu}{\int \tilde{S}_{\text{ref}}(\nu)d\nu}
	\end{aligned}
	\label{eq:P_det}
\end{equation}

\begin{equation}
	\begin{aligned}
		R_{\text{peak,det}}=&U_{\text{LI,det}}/P_{det}
		\\
		=&2R_{\text{peak,ref}}\frac{U_{\text{LI,det}}}{U_{\text{LI,ref}}}\frac{A_{\text{ref}}}{A_{\text{det}}}
		\frac{\int \tilde{S}_{\text{ref}}(\nu)\text{d}\nu}{\int \tilde{S}_{\text{det}}(\nu)\text{d}\nu}
	\end{aligned}
	\label{eq:R_total}
\end{equation}

We used a MCT detector as reference with $R_{\text{peak,ref}}=\SI{5e4}{\volt/\watt}$ to measure the responsivity of our QWIP. We used a transimpedance amplifier on the QWIP with a gain $G=\SI{1e5}{\volt/\ampere}$. We used a $\SI{1000}{\celsius}$ black body source. The detectors were fully covered by the black body radiation. The detection areas of the detectors are respectively $A_{\text{MCT}}=\SI{250}{\micro\meter}\times\SI{250}{\micro\meter}$ and $A_{\text{QWIP}}=\SI{80}{\micro\meter}\times\SI{35}{\micro\meter}$.

We measured $U_{\text{LI,MCT}}=\SI{25.5}{\volt}$ and $U_{\text{LI,QWIP}}=\SI{40}{\nano\volt}$. Numerically, we obtain $\int \tilde{S}_{\text{MCT}}(\nu)d\nu=2845$ and $\int \tilde{S}_{\text{QWIP}}(\nu)d\nu=963.6$. Therefore the received optical power on the QWIP is $P_{det}=\SI{4.2}{\micro\watt}$ and the peak responsivity of the detector is $R_{\text{peak,det}}= \SI{11}{\milli\ampere/\watt}$.

\subsection{Noise levels}

We use a \emph{UHFLI} numerical lock-in amplifier to analyse the noise spectra of the detector. The detector is connected to a $\SI{10}{\kilo\ohm}$ trans-impedance amplifier and to the \emph{UHFLI} instrument which calculates the power spectral density of the amplified signal. The measurement bandwidth $\Delta f$ was set to $\SI{1}{\hertz}$. First, the Johnson-Nyquist noise $\sigma_{i,JN}$ is measured by simply connecting the detector to the instrument without applying any bias and without any incident radiation. Then, a bias is applied to measure the dark-current noise $\sigma_{i,dark}$ plus the Johnson-Nyquist noise. Finally, a laser beam is shined on the detector to measure the photo-current noise $\sigma_{i,ph}$ plus the Johnson-Nyquist noise and the dark-current noise. By substracting the different noise levels, we retrieve the indivdidual noise contributions that are summarized in \cref{tab:noise}. The measured photo-current differs from the theoretical shot-noise of \cref{eq:shot_noise} by $\SI{4}{\decibel}$. This difference might be caused by relative intensity noise from the laser beam. We calculte the specific detectivity $D^*$ of the QWIP using \cref{eq:specific_detectivity} and we obtain $D^*=\SI{2e6}{\centi\meter\hertz^{1/2}\watt^{-1}}$.

\begin{equation}
	\sigma_{i,shot\text{-}noise}^2=4egi\Delta f
	\label{eq:shot_noise}
\end{equation}

\begin{equation}
	D^*=\frac{R_{\text{peak,det}}\sqrt{A_{\text{det}}}\sqrt{\Delta f}}{\sigma_{i,total}}
	\label{eq:specific_detectivity}
\end{equation}

Where:
\begin{itemize}
    \item $i$ is the current flowing in the detector
    \item $e$ is the elementary charge of an electron
    \item $g$ is the photo-conductive give of the detector. It represents the average number of electrons that circulate in the active region when a photon is absorbed
\end{itemize}

\begin{table}[b]
\centering
\begin{tabular}{|l|c|c|}
\hline
\textbf{Type of noise} & \textbf{Theoretical power (dBm)} & \textbf{Measured power (dBm)} \\ \hline
Johnson-Nyquist noise  & $-194$                           & Below detection threshold    \\ \hline
Dark current noise     & $-172$                          & $-171$                        \\ \hline
Photo-current noise    & $-170$                           & $-166$                        \\ \hline
\end{tabular}
\caption{Theoretical and measured noise level for a $\SI{5}{\micro\meter}$ QWIP. The noise measurements where taken with a resolution bandwidth of $\SI{1}{\hertz}$.}
\label{tab:noise}
\end{table}

\subsection{Bandwidth}

\begin{figure}[t]
    \centering
    \begin{minipage}{0.33\textwidth}
        \includegraphics[width=1\linewidth]{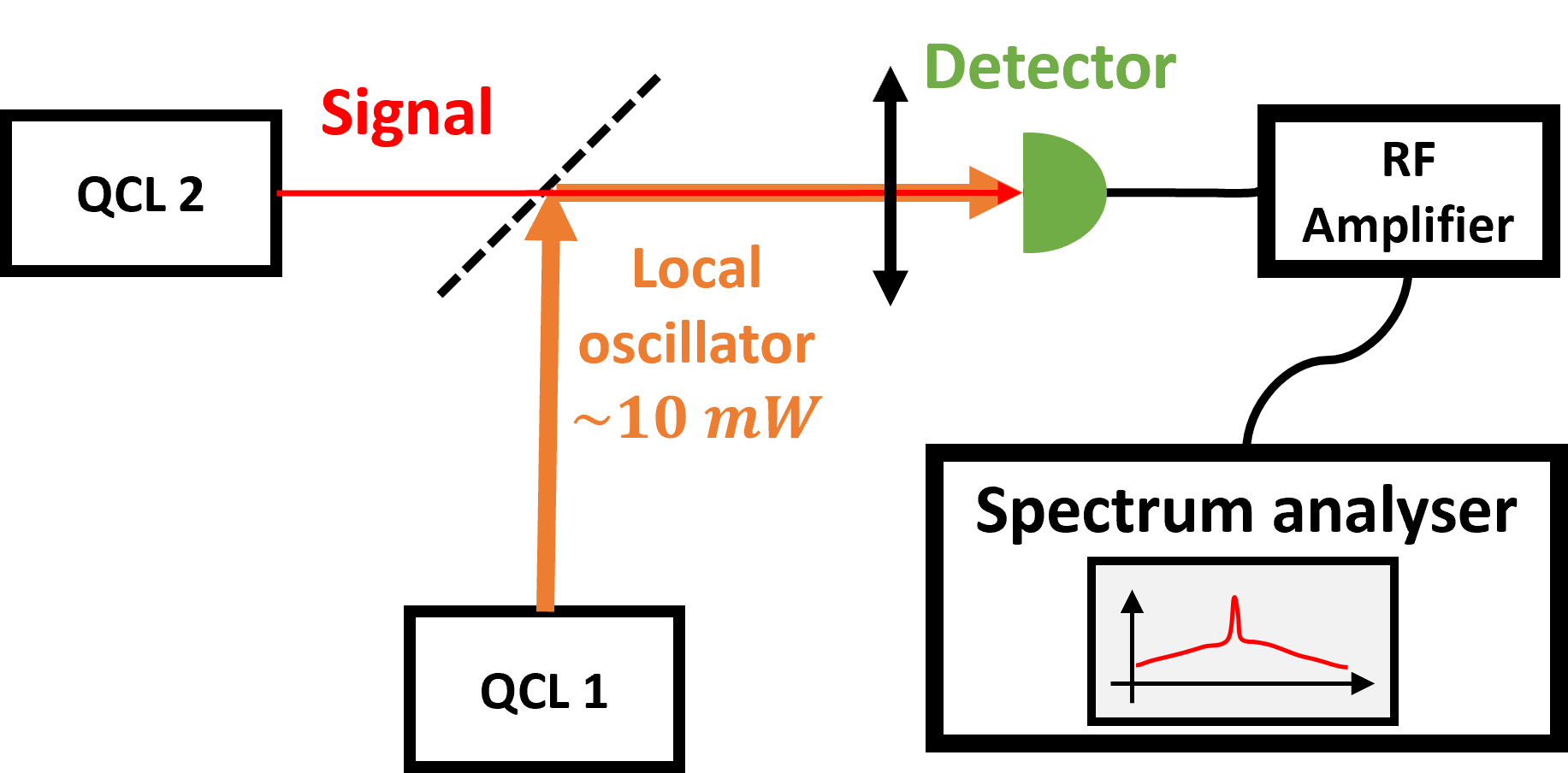}
    \end{minipage}
    \hfill
    \begin{minipage}{0.32\textwidth}
        \includegraphics[width=1\linewidth]{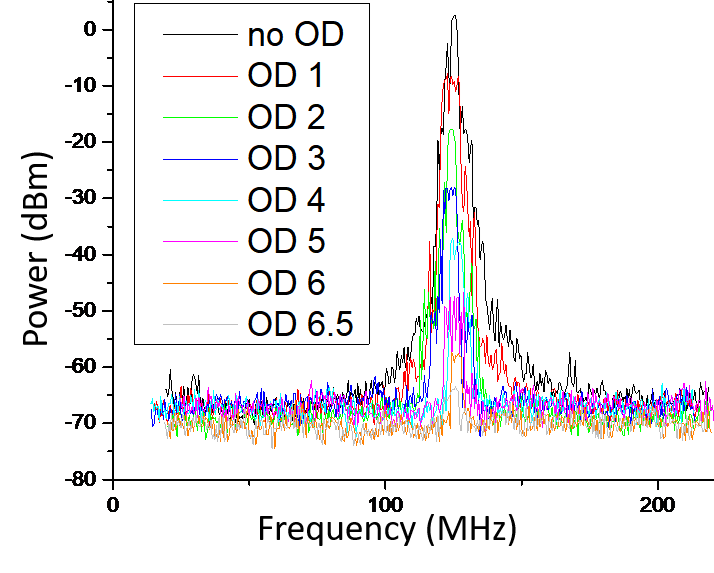}
    \end{minipage}
    \hfill
    \begin{minipage}{0.33\textwidth}
        \includegraphics[width=1\linewidth]{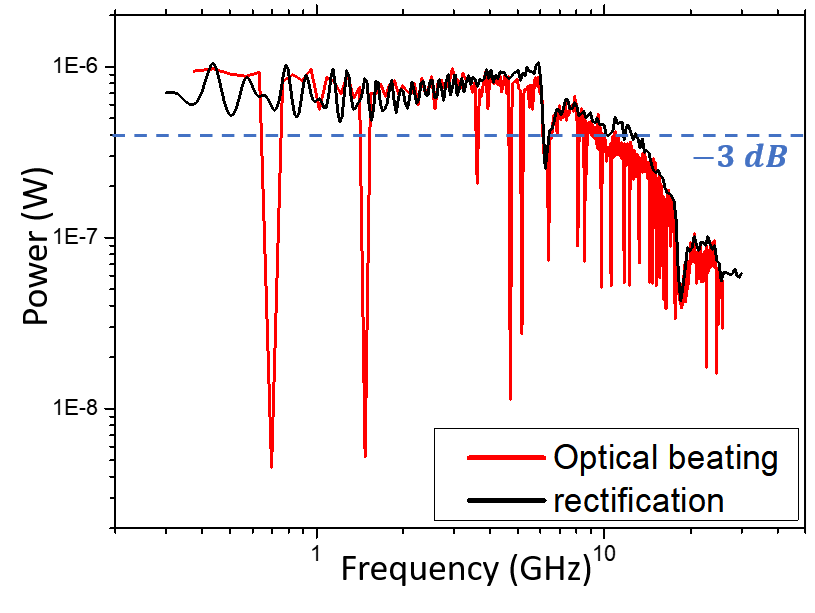}
    \end{minipage}
    \caption{\capleft~Heterodyne set-up for measuring the bandwidth of a detector. \capcenter~Example of heterodyne beatings at $\SI{125}{\mega\hertz}$ viewed on a spectrum analyser with different optical densities (OD). \capright~Measured bandwidth of a $\SI{5}{\micro\meter}$ QWIP.}
    \label{fig:bandwidth}
\end{figure}

The bandwidth of a detector is retrieved using two different methods. The first method consists in electrically sending a sweeping RF signal to the detector and measuring the response of the detector to this signal. The obtained bandwidth for our QWIP using this method is shown on \cref{fig:bandwidth} \captext{Right}. The $\SI{-3}{\decibel}$ frequency bandwidth of this detector is $\SI{10}{\giga\hertz}$.

The second method uses the heterodyne beating between two lasers. The measurement set-up is shown on \textendash~\cref{fig:bandwidth} \capleft. Two QCLs are aligned and focused on the detector to be characterised using a beam splitter. One the QCL plays the role of the local oscillator while the other plays the role of the signal. The heterodyne RF beating signal on the detector to be characterized is sent to an RF amplifier and measured by a spectrum analyser. For each frequency difference $\Delta \nu$, the RF power is measured as shown is \cref{fig:bandwidth} \capcenter. The heterodyne beatnote at $\SI{125}{\mega\hertz}$ can be seen on top of the noise floor. Decreasing the power of the signal laser using optical densities (OD) decreases the power of the heterdoyne beating. The frequency response of the detector is retrieved by sweeping the heterodyne beating frequency between two lasers. The corresponding frequency response of our QWIP is shown on \cref{fig:bandwidth} \captext{Right}. The comparison with the electrical method measurements shows a perfect agreement between the two measurement methods.

\subsection{Two-laser heterodyne detection}

\begin{figure}[]
    \begin{center}
    \includegraphics[width=1\linewidth]{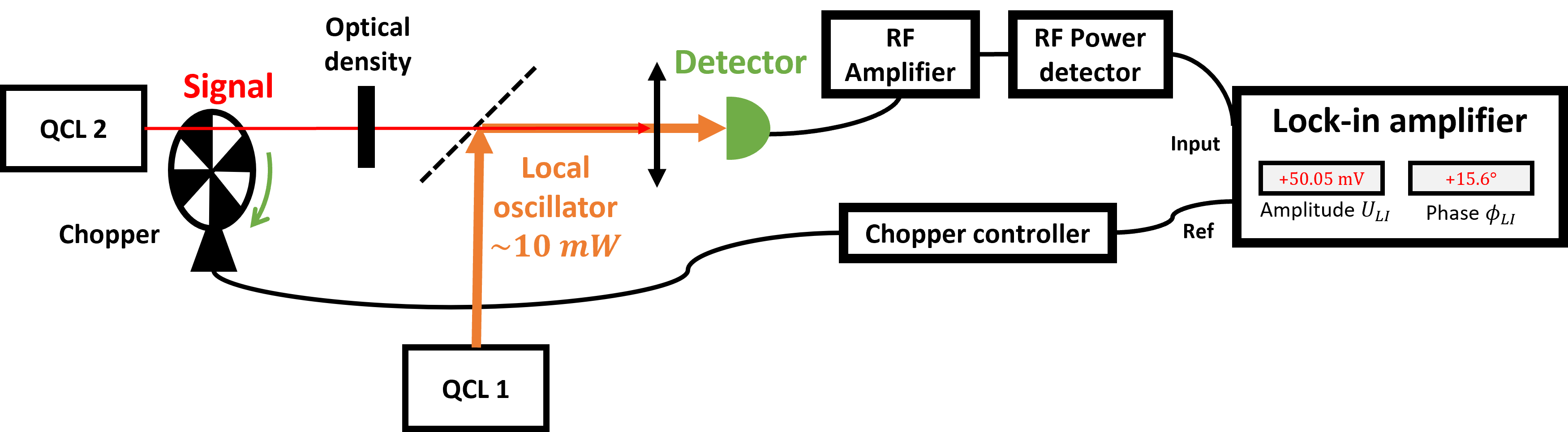}
    \end{center}
    \caption{Two laser heterodyne detection set-up. QCL plays the role of local oscillator while QCL is the signal.}
    \label{fig:QCL_heterodyne}
\end{figure}

Preliminary characterization for heterodyne have been performed using two free running Quantum Cascade Laser. The measurement set-up is described on \cref{fig:QCL_heterodyne}. Two QCLs are aligned and focused on the detector to be characterised using a beam splitter. One the QCL plays the role of the local oscillator while the other plays the role of the signal. The signal beam is attenuated by optical densities and chopped at $\SI{1000}{\hertz}$. The radio frequency signal from the detector is highpass filtered, amplified by a \emph{HD30510} amplifier and sent to a \emph{ZX47-60-LN-S} power meter which outputs a voltage low frequency depending on the RF input power. The output voltage is sent to a \emph{SR830
DSP Lock-In Amplifier} and averaged over a timescale $\tau$.

The local oscillator outputs \SI{41}{\milli\watt} of continuous optical power at $\SI{5}{\micro\meter}$ wavelength, measured before the focusing lens. The signal laser power was \SI{20}{\milli\watt} in the absence of any optical density or chopping. The beating frequency between the two lasers on the QWIP was observed on a spectrum analyser and set to \SI{2}{\giga\hertz} but it was not stabilized. The typical linewidth of the heterodyne beating is $\simeq\SI{10}{\mega\hertz}$. The optical density was gradually increased and measured using the lock-in amplifier. The alignment of the set-up was tuned each time the optical density to maximize the overlap between the signal and the local oscillator on the detector. With a $\tau=\SI{3}{\second}$ characteristic time of integration on the lock-in amplifier, we were able to detect the heterodyne signal with optical densities up to $7.5$ which corresponds to $\frac{1}{2}\times\frac{1}{2}\times\SI{20}{\milli\watt}\times10^{-7.5}\simeq\SI{0.2}{\nano\watt}$ signal power on the detector. The two $1/2$ factors come from the losses due to the signal chopping and the $\SI{45}{\degree}$ facet. This results highlight one of the key property of intersubband detectors which is their very high dynamic range coming from the $\sim\SI{100}{\pico\second}$ quantum processes at stage. Shining a $\SI{40}{\milli\watt}$ local oscillator, which is more than $10^7$ stronger than the signal, does not saturate the detector, thus enabling strong optical amplification in heterodyne detection. This result highlight the very large dynamic range of intersubband detectors.

\section{CONCLUSION AND PERSPECTIVES}
\label{p:patch_antenna}

We described the basics of QCD and QWIP: their working principle, the mesa and the patch-antenna architecture. The timescale of the intersubband phenomena at stage in these detectors enable detection at bandwidth of tens of gigahertz which would improve the signal to noise ratio of heterodyne interferometry in mid infrared. QCD use a set a quantum well and resonnant tunneling to extract the excited electrons without applying any bias and thus exhibit no dark current. The very low noise of QCD, even at ambient temperature, is an asset for detecting low signals with large integration times. QWIP responsivity is higher than those of QCD. However, they exhibit a higher amount of noise, mostly caused by dark current. At cryogenic temperature, the dark current in QWIP vasnishes, making QWIP competitive against QCD. Further improvement can be obtained using meta-materials, namely patch-antenna resonators which couple the electrical field from the light into resonators and increase the bandwidth of the detector.

We explained basic measurement methods that are used to characterize intersubband detectors in terms of responsivity, noise levels and bandwidth. These methods have been used on a $\SI{5}{\micro\meter}$ QWIP at ambiant temperature. We measured a responsivity of $\SI{11}{\milli\ampere/\watt}$, a specific detectivity of $D^*=\SI{2e6}{\centi\meter\hertz^{1/2}\watt^{-1}}$ and a bandwdith of $\SI{10}{\giga\hertz}$. We also measured the minimum detectable signal power on a two laser heterodyne set-up. With free running lasers, we were able to detect the incoming signal down to $\SI{0.2}{\nano\watt}$. Future work will include direct comparison between the QWIP and QCD at $\SI{9}{\micro\meter}$ wavelength and at ambient temperature and at cryogenic temperature. The lower temperature decreases thermal noise while improving the responsivity of the detectors, thus strongly increasing the specific detectivity. In Palaferri 2018\cite{palaferri2018} , specific detectivity of $D^*=\SI{2e10}{\centi\meter\hertz^{1/2}\watt^{-1}}$ was demonstrated at $\SI{9}{\micro\meter}$ on patch-antenna QWIP at cryogenic temperature.

We measured the heterodyne beating between two free-running QCL. In this configuration, we were able to detect the signal down to $\simeq \SI{0.2}{\nano\watt}$ of power. The next step on this work will consist in replacing the coherent laser signal with an incoherent black-body to emulate large bandwidth heterodyne detection on an astronomical object. Then, we plan to use QCDs and QWIPs on a full interferometric heterodyne set-up: see Berger \textit{et al} \emph{SPIE 12183-40}.

The current demonstrations of high detectivity QCD and QWIP with large bandwidths make theses detectors interesting for astronomical applications in the mid-infrared. The potential of future improvements to reduce noise while increasing the detectivity and the bandwidth of QCDs and QWIPS using meta-materials and faster electronics reinforces our interest in studying these devices from an astronomical point of view.

\acknowledgments 

We acknowledge financial support from \emph{ENS-THALES Chair} and from \emph{LabEx FOCUS ANR-11-LABX-0013}.

\newpage
\bibliography{main} 

\begin{thebibliography}{1}

\bibitem{monnier2018}
Monnier, J.~D., Kraus, S., Ireland, M.~J., Baron, F., Bayo, A., Berger, J.-P.,
  Creech-Eakman, M., Dong, R., Duch{\^{e}}ne, G., Espaillat, C., Haniff, C.,
  Hönig, S., Isella, A., Juhasz, A., Labadie, L., Lacour, S., Leifer, S.,
  Merand, A., Michael, E., Minardi, S., Mordasini, C., Mozurkewich, D.,
  Olofsson, J., Paladini, C., Petrov, R., Pott, J.-U., Ridgway, S., Rinehart,
  S., Stassun, K., Surdej, J., ten Brummelaar, T., Turner, N., Tuthill, P.,
  Vahala, K., van Belle, G., Vasisht, G., Wishnow, E., Young, J., and Zhu, Z.,
  ``The planet formation imager,'' {\em Experimental Astronomy}~{\bf 46},
  517--529 (jul 2018).

\bibitem{ireland2014}
Ireland, M.~J. and Monnier, J.~D., ``{A dispersed heterodyne design for the
  planet formation imager},'' in [{\em Optical and Infrared Interferometry
  IV}{\nolinebreak\hspace{0.1em}]},  Rajagopal, J.~K., Creech-Eakman, M.~J.,
  and Malbet, F., eds.,  {\bf 9146},  339 -- 346, International Society for
  Optics and Photonics, SPIE (2014).

\bibitem{bourdarot2022}
Bourdarot, G., Berger, J.-P., and de~Chatellus, H.~G., ``Multi-delay photonic
  correlator for wideband rf signal processing,'' {\em Optica}~{\bf 9},
  325--334 (Apr 2022).

\bibitem{palaferri2018}
Palaferri, D., Todorov, Y., Bigioli, A., Mottaghizadeh, A., Gacemi, D.,
  Calabrese, A., Vasanelli, A., Li, L., Davies, A.~G., Linfield, E.~H.,
  Kapsalidis, F., Beck, M., Faist, J., and Sirtori, C., ``Room-temperature
  nine-$\mathrm{\mu}$m-wavelength photodetectors and {GHz}-frequency heterodyne
  receivers,'' {\em Nature}~{\bf 556},  85--88 (Mar. 2018).

\bibitem{steinkogler2003}
Steinkogler, S., Schneider, H., Walther, M., and Koidl, P., ``Determination of
  the electron capture time in quantum-well infrared photodetectors using
  time-resolved photocurrent measurements,'' {\em Applied Physics Letters}~{\bf
  82}(22),  3925--3927 (2003).

\bibitem{hakl2021}
Hakl, M., Lin, Q., Lepillet, S., Billet, M., Lampin, J.-F., Pirotta, S.,
  Colombelli, R., Wan, W., Cao, J.~C., Li, H., Peytavit, E., and Barbieri, S.,
  ``Ultrafast quantum-well photodetectors operating at 10 $\mu m$ with a flat
  frequency response up to 70 ghz at room temperature,'' {\em ACS
  Photonics}~{\bf 8}(2),  464--471 (2021).

\bibitem{capasso1999}
Capasso,  [{\em Intersubband Transitions in Quantum Wells, Physics and Device
  Applications I}{\nolinebreak\hspace{0.1em}]}, vol.~62 (1999).

\bibitem{rodriguez2022}
Rodriguez, E., Bonazzi, T., Dely, H., Mastrangelo, M., Pantzas, K., Beaudoin,
  G., Sagnes, I., Vasanelli, A., Todorov, Y., and Sirtori, C., ``Metamaterial
  engineering for optimized photon absorption in unipolar quantum devices,''
  {\em Optics Express}~{\bf 30},  20515 (May 2022).

\end{thebibliography}
\bibliographystyle{spiebib} 

\end{document}